\documentclass[aps,prd,twocolumn,eqsecnum,showpacs,amsmath,amssymb,superscriptaddress,nofootinbib]{revtex4-2}

\usepackage[utf8]{inputenc}
\usepackage[english]{babel}
\usepackage{amsmath,amssymb}
\usepackage{amsfonts}
\usepackage{graphicx}
\usepackage{hyperref}
\usepackage{xcolor}
\usepackage{caption}
\definecolor{linkcolor}{named}{black}
\hypersetup{ linkcolor=linkcolor, colorlinks=true}
 
\newcommand{\beq}{\begin{equation}}
\newcommand{\eeq}{\end{equation}}
\newcommand{\bea}{\begin{eqnarray}}
\newcommand{\eea}{\end{eqnarray}}

\begin{document}
\title{Image of an accreting general Ellis-Bronnikov wormhole}

\author{Valeria A. Ishkaeva}
\email{ishkaeva.valeria@mail.ru}
\affiliation{Institute of Physics, Kazan Federal University, Kremliovskaya str. 16a, Kazan 420008, Russia}

\author{Sergey V. Sushkov}
\email{sergey$_\,$sushkov@mail.ru}
\affiliation{Institute of Physics, Kazan Federal University, Kremliovskaya str. 16a, Kazan 420008, Russia}

\date{\today}

\begin{abstract}
Gravitational lensing properties of supermassive astrophysical objects, such as black holes and wormholes, provide the realistic way for their discovering and investigating. Various lensing effects in a wormhole spacetime have been widely studied in the literature. One of the most popular object for investigation is the Ellis wormhole which represents the simplest wormhole geometry. The Ellis solution represents only the particular case of a general wormhole solution found independently by Ellis 
and Bronnikov. 
Surprisingly but gravitational lensing properties of general Ellis-Bronnikov wormholes are practically not investigated. In this paper we explore in details the propagation of light, forming a shadow and silhouette, and forming an image of accretion disk in the spacetime of the Ellis-Bronnikov wormhole. As well we compare characteristics of images obtained for the Ellis-Bronnikov wormhole with those for the Schwarzschild black hole. This comparison could be useful for future observations of supermassive astrophysical objects.
\end{abstract}

\maketitle

\section{Introduction}
Usually wormholes are considered as tunnels in spacetime with relatively narrow throats that connect two different regions of the same universe, or two different universes. The possible existence of such configurations as
solutions of the gravitational field equations was first mentioned in 
\cite{Flamm, EinsteinRosen, Wheeler_Geons, Wheeler_GDs}. In 70th-80th the only several works dedicated to wormholes were published \cite{Bronnikov, Ellis, Ellis:79, Clement:84a, Clement:84b}. Among of them the first exact traversable wormhole solutions were discussed in \cite{Bronnikov, Ellis} (1973) in the Einstein-scalar theory in which the scalar is of phantom nature, i.e., has a wrong sign of the kinetic term in the Lagrangian.
However, interest in these objects has increased tremendously after the work of M. Morris and K. Thorne \cite{MorrisThorne1} (1988) where the theoretical prospects of use of traversable wormholes for interstellar travels were discussed, and it was shown that, in the framework of Einstein gravity, maintaining a static wormhole throat needs ``exotic'' matter that violates the null energy condition (NEC). Today, the literature devoted to various aspects of wormhole physics is very extensive. For example, overviews of wormhole research can be found in \cite{VisserBook, LoboReview}.

Though nowadays wormholes are rather well studied theoretically, they remain to be hypothetical objects not discovered still by astrophysical methods. One of the most realistic way to discover wormholes as astrophysical objects is to use their gravitational lensing properties \cite{TsuHarYaj}. From an astrophysical point of view, wormholes are massive objects that bend the trajectories of photons passing by them. As the result, wormholes might play a role of gravitational lenses forming Einstein circles and arcs as images of distant galaxies. As well, one might consider wormholes instead of black holes as candidates for supermassive objects in centers of galaxies with active nuclei. Similarly to a black hole spacetime, in a wormhole spacetime there exists a photon sphere forming a so-called shadow which can be observed in modern \cite{EHT} and future observations.   

In last years gravitational lensing properties of wormholes are of great interest and intensively investigating. 
In particular, the propagation of photons, particles and fields in different wormhole geometries had been considered in Refs. 
\cite{
KonoplyaZhidenko, 
SarbachZannias, 
Taylor, 
MisCha:17, 
WilGruKleKun:18,
PotTchTsi:20, 
Deligianni_etal:2021, 
Benavides_etal}.
In our knowledge, the first work where a lensing effect in a wormhole spacetime have been discussed was that by L. Chetouani and G. Cl\'{e}ment \cite{ChetouaniClement}. Later, problems of weak and strong gravitational lensing in wormhole geometries were studied in numerous works 
\cite{Perlick, 
NanZhaZak, 
Muller, 
NakajimaAsada, 
YooHaradaTsukamoto:2013,  
Tsukamoto, 
TsukamotoHarada, 
JusufiOvgun:2018,
Lukmanova_etal:2018,  
Bronnikov:2018nub,
LiHeZhou:2020,
Tsukamoto:2020}. 
Shadows of wormholes were studying in Refs. 
\cite{
Ohgami:2015nra, 
Shaikh:2018oul, 
Wang:2020emr, 
Wielgus_etal:2020, 
PerlickTsupko, 
Bugaev:2021dna, 
Bronnikov:2021liv, 
Jusufi:2021lei, 
Sokoliuk:2022owk} 
for static spherically symmetric configurations, and in Refs. 
\cite{
Nedkova:2013msa, 
OhgamiSakai:2016, 
Shaikh:2018kfv, 
Amir:2018pcu, 
Kasuya:2021cpk, 
Rahaman:2021web} 
for rotating wormholes. Additionally, the recent review of past and current efforts to search for astrophysical wormholes in the Universe can be found in \cite{BambiStojkovic}.

It is necessary to emphasize that a wormhole configuration is not vacuum, and hence there are no preferred wormhole geometries, such as Schwarzschild or Kerr ones for black holes. For this reason, numerous examples of different wormhole models have been considered in the literature. One of the most popular object for investigation is the Ellis wormhole, which has been studied in Refs. 
\cite{
TsuHarYaj, 
SarbachZannias, 
Ohgami:2015nra, 
PerlickTsupko, 
Bugaev:2021dna, 
OhgamiSakai:2016,	
Perlick, 
NanZhaZak, 
Muller, 
YooHaradaTsukamoto:2013,
Tsukamoto, 
TsukamotoHarada,
Lukmanova_etal:2018,
Bronnikov:2018nub}.
The Ellis wormhole spacetime is a static spherically symmetric solution to the Einstein equations with a massless, minimally coupled phantom scalar field \cite{Bronnikov, Ellis} with the metric
\begin{equation} 
	\label{Ellis_wh}
	ds^2=-dt^2+dr^2+(r^2+a^2) (d\theta^2+\sin{\theta}^2d\varphi^2),
\end{equation}
where $a$ is an arbitrary parameter corresponding to the radius of the wormhole throat located at $r=0$. Note that the Ellis geometry is symmetric with respect to the throat, $r\leftrightarrow -r$. Moreover, since $g_{00}=const=-1$, the Ellis wormhole is asymptotically massless. Here we would like to remind a reader that the metric (\ref{Ellis_wh}) represents only the particular case of a general wormhole solution found independently by Ellis \cite{Ellis} and Bronnikov \cite{Bronnikov}. Generally, the Ellis-Bronnikov wormhole\footnote{
	Bronnikov himself has suggested to use the term ``anti-Fisher"  in order to recall that the corresponding solution for a canonical scalar field was first obtained by I.Z. Fisher in 1948 \cite{Fisher}. However, we prefer to coin this solution as the Ellis-Bronnikov wormhole.} 
is not symmetric with respect to the throat and has nonzero asymptotic mass. Surprisingly but gravitational lensing properties of general Ellis-Bronnikov wormholes are practically not investigated. We can point out the only paper \cite{Bronnikov:2018nub} where some lensing features of asymmetric wormholes and, in particular, the Ellis-Bronnikov wormhole were discussed. 

{\em {\bf Notice:} Just after publishing our manuscript in arXiv [arXiv:2308.02268 [gr-qc]], we received a message from colleagues who drawn our attention to the papers \cite{Cai:2023ite, Huang:2023yqd}, which we missed. In Ref. \cite{Cai:2023ite}, the authors investigate the gravitational lensing effect at higher order under weak-field approximation in the Ellis-Bronnikov wormhole spacetime. The second work \cite{Huang:2023yqd} is more closely related to our investigation. As well as in our work, the authors of Ref. \cite{Huang:2023yqd} calculated the geodesics and classified the photon trajectories in the Ellis-Bronnikov wormhole spacetime, obtained the location of the photon sphere and the corresponding value of impact parameter, found the innermost stable circular orbit (ISCO). Though some results of our work are particularly coincided with those in \cite{Huang:2023yqd}, our work contains a number of novel results comparing with Ref. \cite{Huang:2023yqd}.}

As we will demonstrate later, the geometry of the general Ellis-Bronnikov wormhole has essential differences in comparison with the particular case --- the Ellis wormhole. 
In this paper we explore in details the propagation of light, forming a shadow and silhouette, and forming an image of accretion disk in the spacetime of the Ellis-Bronnikov wormhole. As well we compare characteristics of images obtained for the Ellis-Bronnikov wormhole with those for the Schwarzschild black hole.

The article is organized as follows. In Section \ref{Sec_EBWH}, we introduce the metric of the Ellis-Bronnikov wormhole spacetime and discuss the features of its geometry. In Section \ref{Sec_PT}, we obtain the equations of motion of particles in the Ellis-Bronnikov wormhole spacetime and derive the position of the circular orbit of photons and the position of the innermost stable circular orbit of particles. In Section \ref{Sec_WO}, we describe the difference between a wormhole shadow and its throat silhouette, derive their sizes for the Ellis-Bronnikov wormhole, and compare the results with those obtained for the sizes of the Schwarzschild black hole shadow and its event horizon silhouette. And in Section 4, we obtain an expression for the energy shift of photons reaching the observer, build images of the accreting Ellis-Bronnikov wormhole and the accreting Schwarzschild black hole, and compare them with each other.

\section{Ellis-Bronnikov wormhole} \label{Sec_EBWH}
\subsection{Metric}
The metric of the Ellis-Bronnikov wormhole spacetime can be cost in the following form: 
\begin{equation}\label{BEmetric}
    ds^2=-e^{2u(r)}dt^2+e^{-2u(r)}\left[dr^2+(r^2+a^2) d\Omega^2\right],
\end{equation}
where $d\Omega^2=d\theta^2+\sin^2{\theta}d\varphi^2$, $r$ is the radial coordinate running from $-\infty$ to $+\infty$, $m$ and $a$ are two free parameters, and
\begin{equation}
    u(r)=\frac{m}{a}\left(\arctan{\frac{r}{a}}-\frac{\pi}{2}\right).
\end{equation}

Taking into account the following asymptotic behavior:
\bea
\left.e^{2u(r)}\right|_{r\to+\infty} &=& \left(1-\frac{2m}{r}\right)+O\left(r^{-2}\right), 
\nonumber\\
\left.e^{2u(r)}\right|_{r\to-\infty} &=&
e^{-\frac{2\pi m}{a}}\left(1+\frac{2m}{|\,r\,|}\right)+O\left(|\,r\,|^{-2}\right),
\nonumber
\eea
one may see that the spacetime with the metric (\ref{BEmetric})
possesses by two asymptotically flat regions ${\cal R}_\pm: r\to\pm\infty$. The asymptotical mass in the region ${\cal R}_{+}$ is equal to $m_+=m$, while  in the region ${\cal R}_{-}$ is $m_{-} = - m\, \exp(-{\pi m}/{a})$.
Note that the masses have both different values and different signs. Therefore, the Bronnikov-Ellis wormhole appears as an object with a negative mass for a distant observer in the region ${\cal R}_{-}$. The asymptotical regions ${\cal R}_{+}$ and ${\cal R}_{-}$ are connected by the throat whose radius corresponds to the minimum of the radius of two-dimensional sphere, $R^2(r)=e^{-2u(r)}(r^2+a^2)$. The minimum of $R(r)$ is achieved at $r_{th} = m$ and equal to
\begin{equation}
    R_0=\exp\left[-\frac{m}{a}\left(\arctan{\frac{m}{a}}-\frac{\pi}{2}\right)\right]\left(m^2+a^2\right)^{1/2}.
\end{equation}
Note that in case $m=0$ one has $m_\pm=0$, and the metric (\ref{BEmetric}) reduces to the Ellis metric (\ref{Ellis_wh}).

\subsection{Embedding diagram}
\begin{figure}[t!]
	\begin{minipage}[h]{0.47\linewidth}
		\center{\includegraphics[width=1\linewidth]{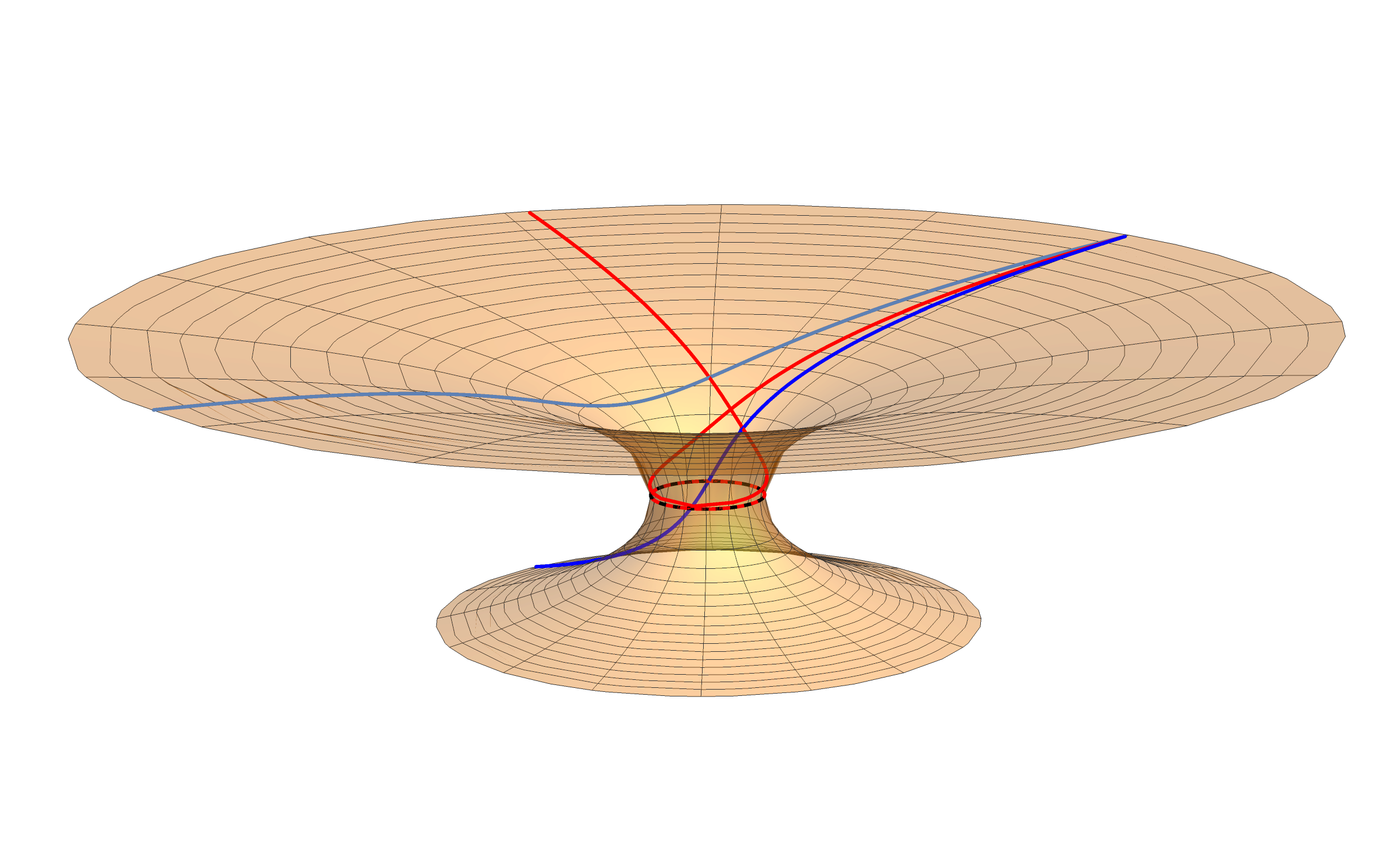}}
	\end{minipage}
	\hfill
	\begin{minipage}[h]{0.47\linewidth}
		\center{\includegraphics[width=1\linewidth]{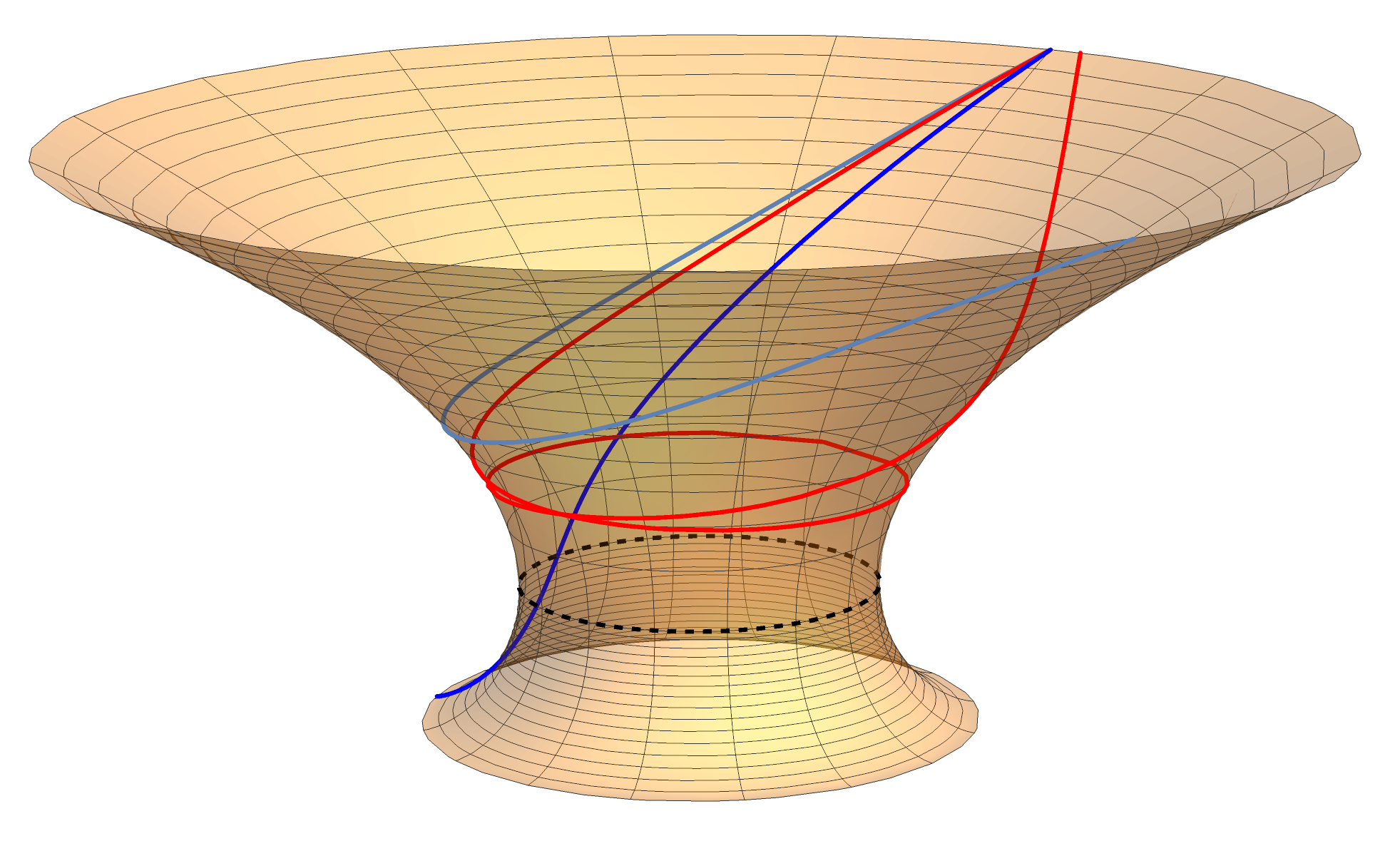}}
	\end{minipage}
	\caption{\label{emdiag} Embedding diagrams for the Ellis-Bronnikov wormhole at $m=0, a=1$ (left) and at $m=1, a=1$ (right). The red curves are trajectories of photons moving along cyclic orbits; the black dotted circle is the throat.}
\end{figure}

An embedding diagram is the standard way to visualize a wormhole geometry and geodesic motion of both massive and massless test particles in this geometry.
To construct the embedding diagram for the Ellis-Bronnikov wormhole (\ref{BEmetric}) we consider its two-dimensional section $t=const$ (surface of constant time) and $\theta=\pi/2$ (equatorial plane):  
\begin{equation}\label{ds1}
	ds^2_{(2)}=e^{-2u(r)}dr^2+e^{-2u(r)}\left(r^2+a^2\right)d\varphi^2.
\end{equation}
Then, as usually, we suppose that this surface is embedded into three-dimensional euclidean space with the metric given in cylindrical coordinates:
\begin{equation}
	ds^2= dz^2+dR^2+R^2 d\varphi^2.
\end{equation}
An embedded axially symmetric surface is determined as $z=z(R)$, where $z(R)$ is the embedding function. A metric induced on the embedded surface reads
\begin{equation}\label{ds2}
	ds^2= (z'^2+1)dR^2+R^2 d\varphi^2,
\end{equation}
where $z'=dz/dR$. Comparing the metrics (\ref{ds1}) and (\ref{ds2}), one can find
\begin{eqnarray}
	R^2(r) &=& e^{-2u(r)}\left(r^2+a^2\right),
\\
\label{em2}
	z'^2 &=& \frac{a^2 - m^2 +2mr}{(r-m)^2} \,.
\end{eqnarray}
Since $z'^2$ is not negative, from Eq. \Ref{em2} we have the constraint $a^2 - m^2 + 2mr \ge 0$, i.e.   
\beq
r\geq r_0 \,,
\eeq
where $r_0=(m^2-a^2)/{2m}$. Thus, the embedding diagram is determined within the interval $r\in[r_0,\infty)$. Note also that $r_0\to-\infty$ in case $m\to 0$. Examples of embedding diagrams are shown in Fig. \ref{emdiag}.

\section{Particle trajectories} \label{Sec_PT}
\subsection{Equations of motion for test particles}
Geodesic equations can be derived from the Hamilton-Jacobi equation 
\begin{equation}
	\mu^2=-g^{i j}\frac{\partial S}{\partial x^{i}}\frac{\partial S}{\partial x^{j}},\label{HY}
\end{equation}
where $S=S(x^{i})$ is the Jacobi action, $g^{i j}$ are components of the metric and $\mu$ is the rest mass of the particle ($\mu=0$ for photons). Taking into account the cyclicity of $t$ and $\varphi$, one can write the action $S$ as follows \cite{Carter}:
\begin{equation}
	S=-Et+L\varphi+S_r(r)+S_\theta(\theta),\label{S}
\end{equation}
where $E\equiv-p_t$ is the total energy and $L\equiv~p_{\varphi}$ is the azimuthal angular momentum. 
Thus, from equations (\ref{HY}) and (\ref{S})
\begin{widetext}
\begin{equation}
	-(r^2+a^2)\left(\frac{dS_r}{dr}\right)^2+e^{-4u(r)}(r^2+a^2)E^2-e^{-2u(r)}(r^2+a^2)\mu^2=\left(\frac{dS_\theta}{d\theta}\right)^2+\frac{L^2}{\sin^2\theta}
	\equiv K,
\end{equation}
\end{widetext}
where $K$ is a constant of integration.
Using the equality
\begin{equation}
	\frac{dx^{i}}{d\lambda}=g^{ij}\frac{\partial S}{\partial x^{j}},
\end{equation}
we can write the equations of motion in the following form
\begin{eqnarray}
	\frac{dt}{d\lambda}&=&E\cdot e^{-2u(r)},\label{ut}\\
	\frac{dr}{d\lambda}&=&\pm\sqrt{E^2-\frac{e^{4u(r)}}{r^2+a^2}(Q+L^2)-\frac{\mu^2}{E^2}e^{2u(r)}} 
	\nonumber\\
	&=& \pm\sqrt{R(r)},\label{R}\\
	\frac{d\theta}{d\lambda}&=&\pm\frac{e^{2u(r)}}{r^2+a^2} \sqrt{Q-L^2\cot^2\theta}
	\nonumber\\
	&=&\pm\frac{e^{2u(r)}}{r^2+a^2}\sqrt{\Theta(\theta)},\label{theta}\\
	\frac{d\varphi}{d\lambda}&=&\frac{L\cdot e^{2u(r)}}{\left(r^2+a^2\right)\sin^2\theta},\label{uphi}
\end{eqnarray} 
where $\lambda$ is related to the particle’s proper time by $\tau/\mu$ and is an affine parameter in the case $\mu\to0$, $Q=K-L^2$ is the Carter constant related to the nonazimuthal part of the particle angular momentum. 
Note that the trajectories of massive particles $(\mu\neq0)$ are determined by three impact parameters: $\gamma=E/\mu$, $l = L/E$ and $q = \sqrt{Q}/E$. In the case of $\mu=0$, the third term in (\ref{R}) disappears, i.e. photon trajectories are determined by two parameters: $l = L/E$ and $q = \sqrt{Q}/E$.

\subsection{Photon trajectories and circular orbits}
In this section we consider trajectories and circular orbits of photons supposing $\mu=0$. 
Analyzing the equations of motion of photons, one can come to the conclusion that there are three types of photon trajectories: (i) trajectories that have a turning point ($R(r)=0$ at some point of the trajectory), go around the wormhole and go to infinity; (ii) trajectories that do not have a turning point ($R(r)>0$ for any point of the trajectory) cross the wormhole throat and go to another region of space; and (iii) trajectories infinitely twisting around the wormhole ($R(r)=0, R'(r)=0$ \cite{BardeenPress}), which move along so-called circular orbits. Thus, circular orbits of photons are those separating the other two types of photon trajectories. 

As we will see later, photons that move in circular orbits form the boundary of the wormhole shadow, so it is important to find the positions of the circular orbit $r_{ph}$ and the impact parameter of the photons $b^2_{ph}=l^2+q^2$ that move along it. Since the space-time of the Ellis-Bronnikov wormhole is spherically symmetric, without loss of generality we will assume that photons move in the equatorial plane $\theta = \pi/2$, hence the parameter $q=0$ and $b_{ph }=l$.
Circular orbits of photons are given by the following conditions \cite{BardeenPress}
\begin{equation}
	\left. \frac{dr}{d\lambda} \right|_{r=r_{ph}}=0,\quad \left. \frac{d^2r}{d\lambda^2} \right|_{r=r_{ph}}=0.
\end{equation}
Then, using the equation (\ref{R}), the position of the circular orbit $r_{ph}$ and the impact parameter of photons moving along these orbits $b_{ph}$ can be expressed in terms of the wormhole parameters:
\begin{equation}
	r_{ph}=2m,\quad  |b_{ph}|=e^{\frac{-2m}{a}\left(\arctan\frac{2m}{a}-\frac{\pi}{2}\right)}\sqrt{4m^2+a^2}.
\end{equation}

\begin{figure}[t!]
	\begin{minipage}[h]{0.325\linewidth}
		\includegraphics[width=1\linewidth]{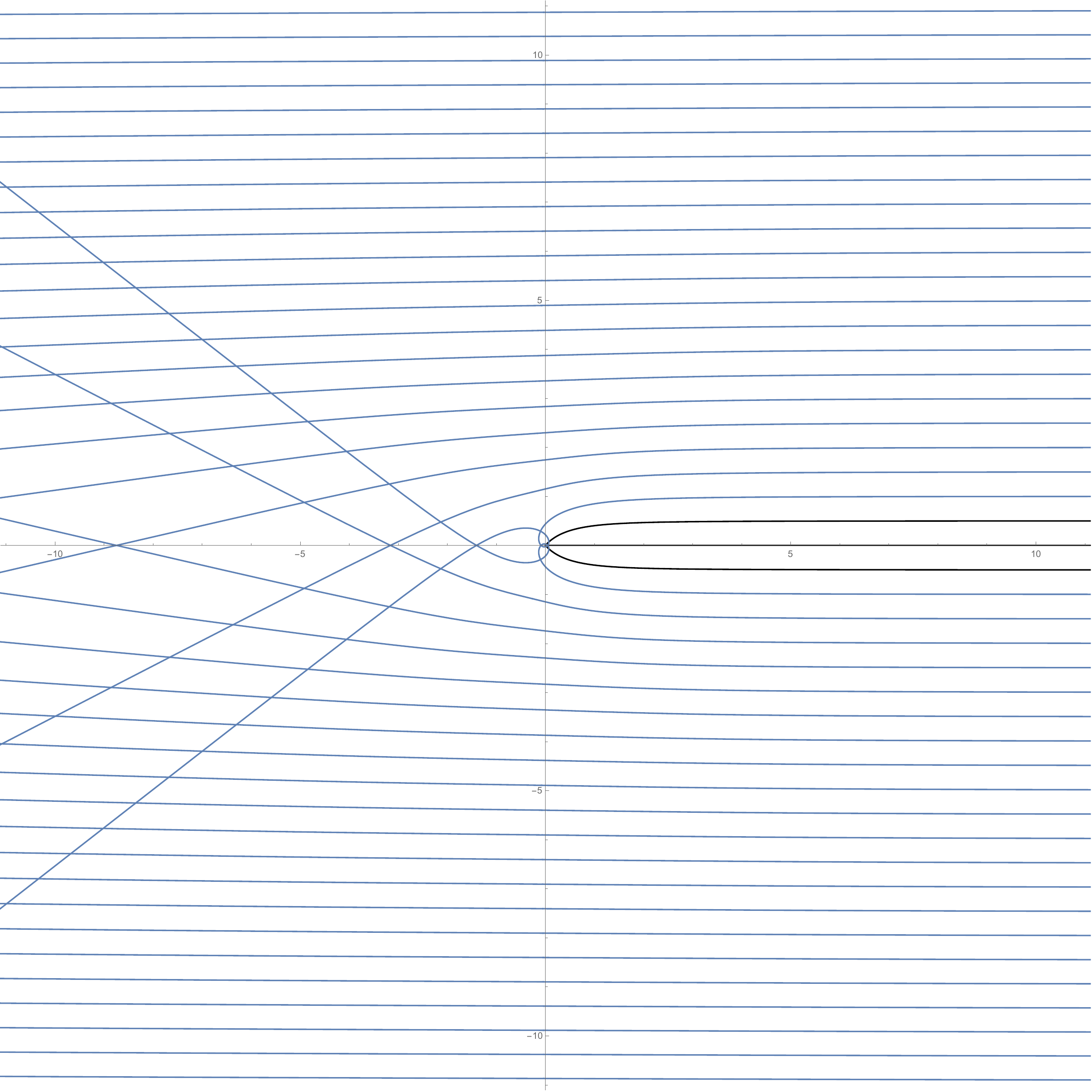}
		\caption*{\scriptsize{$m=0$, $a=1$, $r_{ph}=0$, $|b_{ph}|=1$}}
	\end{minipage}
	\hfill
	\begin{minipage}[h]{0.325\linewidth}
		\includegraphics[width=1\linewidth]{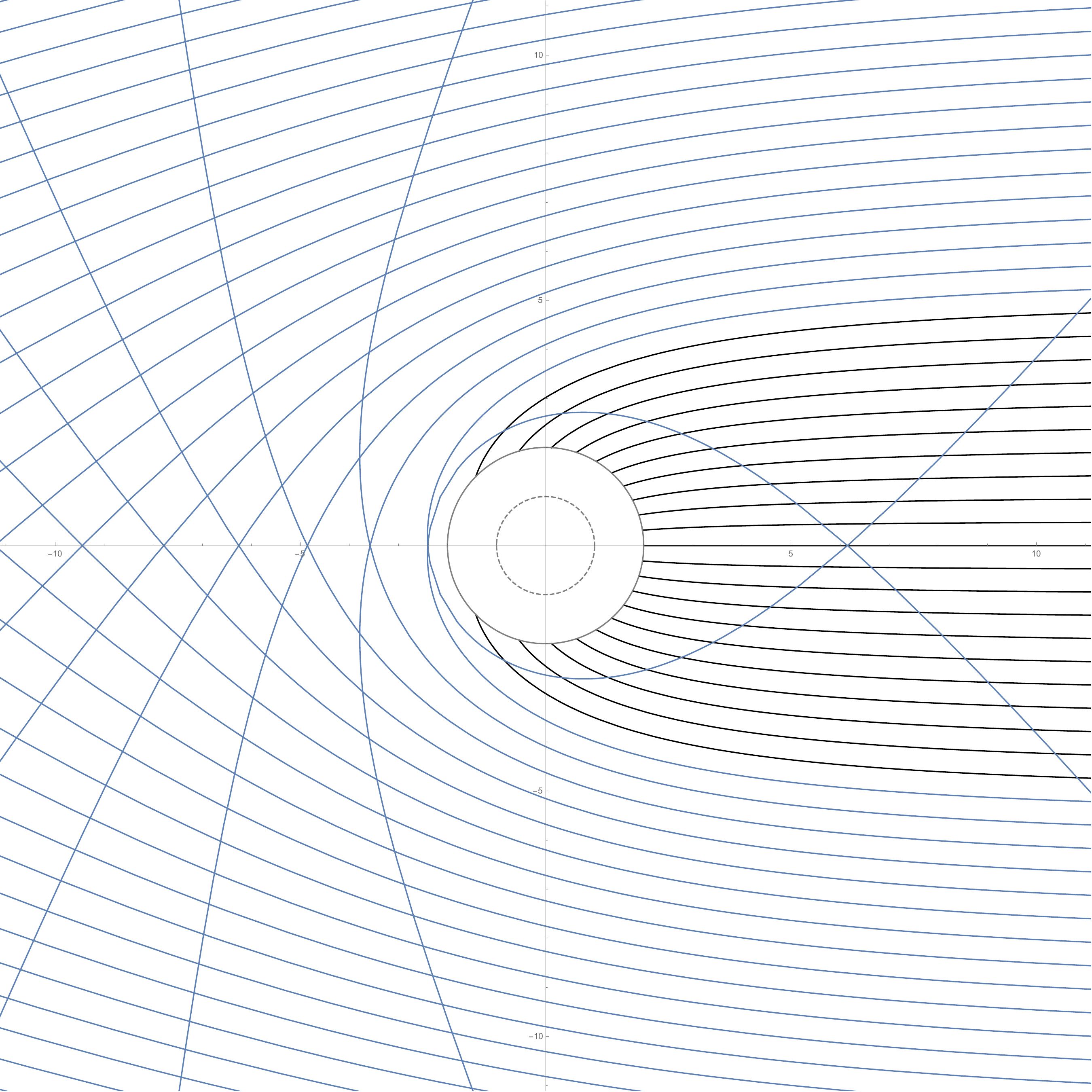}
		\caption*{\scriptsize{$m=1$, $a=1$, $r_{ph}=2$, $|b_{ph}|\approx5.6520$}}
	\end{minipage}
	\hfill
	\begin{minipage}[h]{0.325\linewidth}
		\includegraphics[width=1\linewidth]{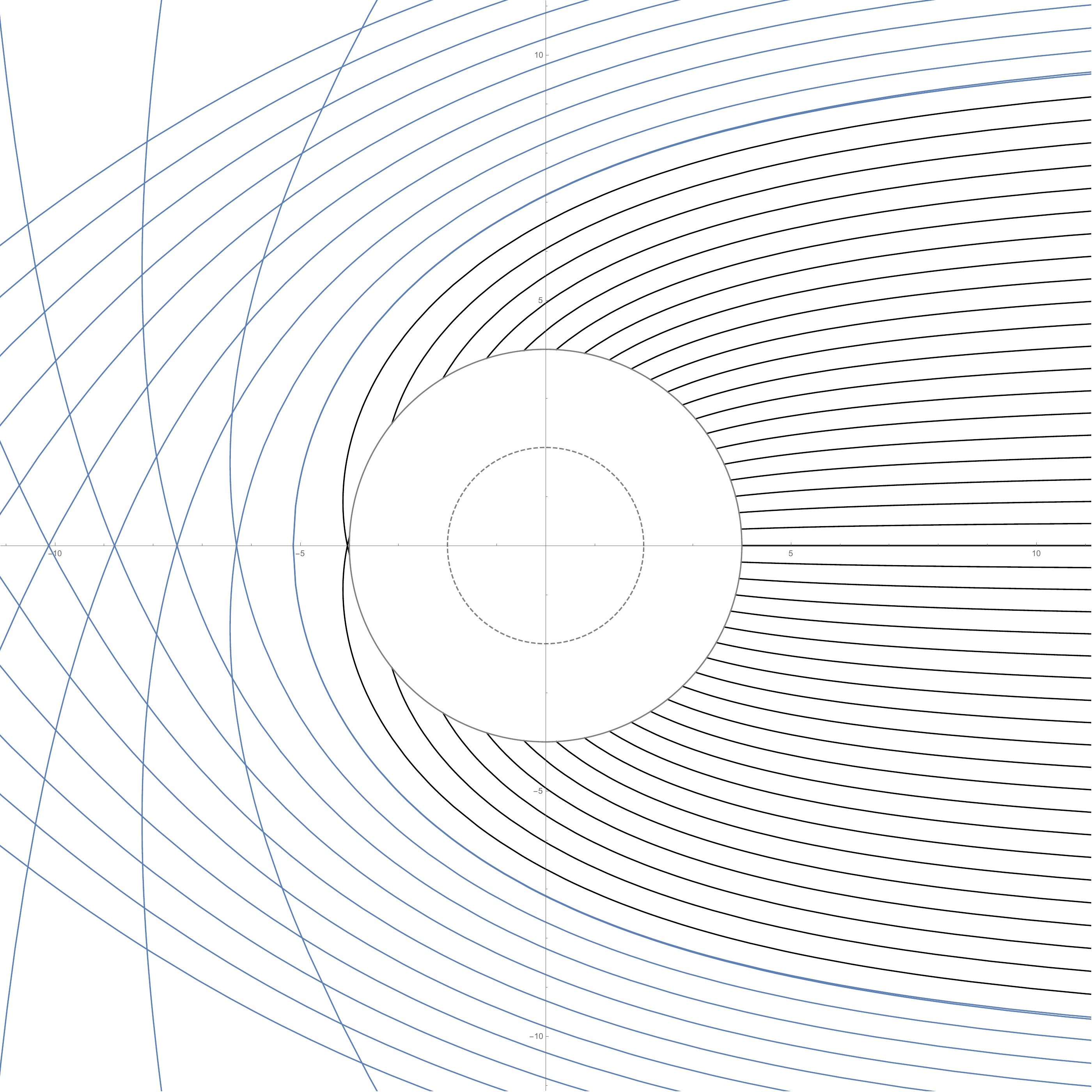}
		\caption*{\scriptsize{$m=2$, $a=1$, $r_{ph}=4$, $|b_{ph}|\approx10.9849$}}
	\end{minipage}
	\vspace{0.02\linewidth}
	\begin{minipage}[h]{0.325\linewidth}
		\includegraphics[width=1\linewidth]{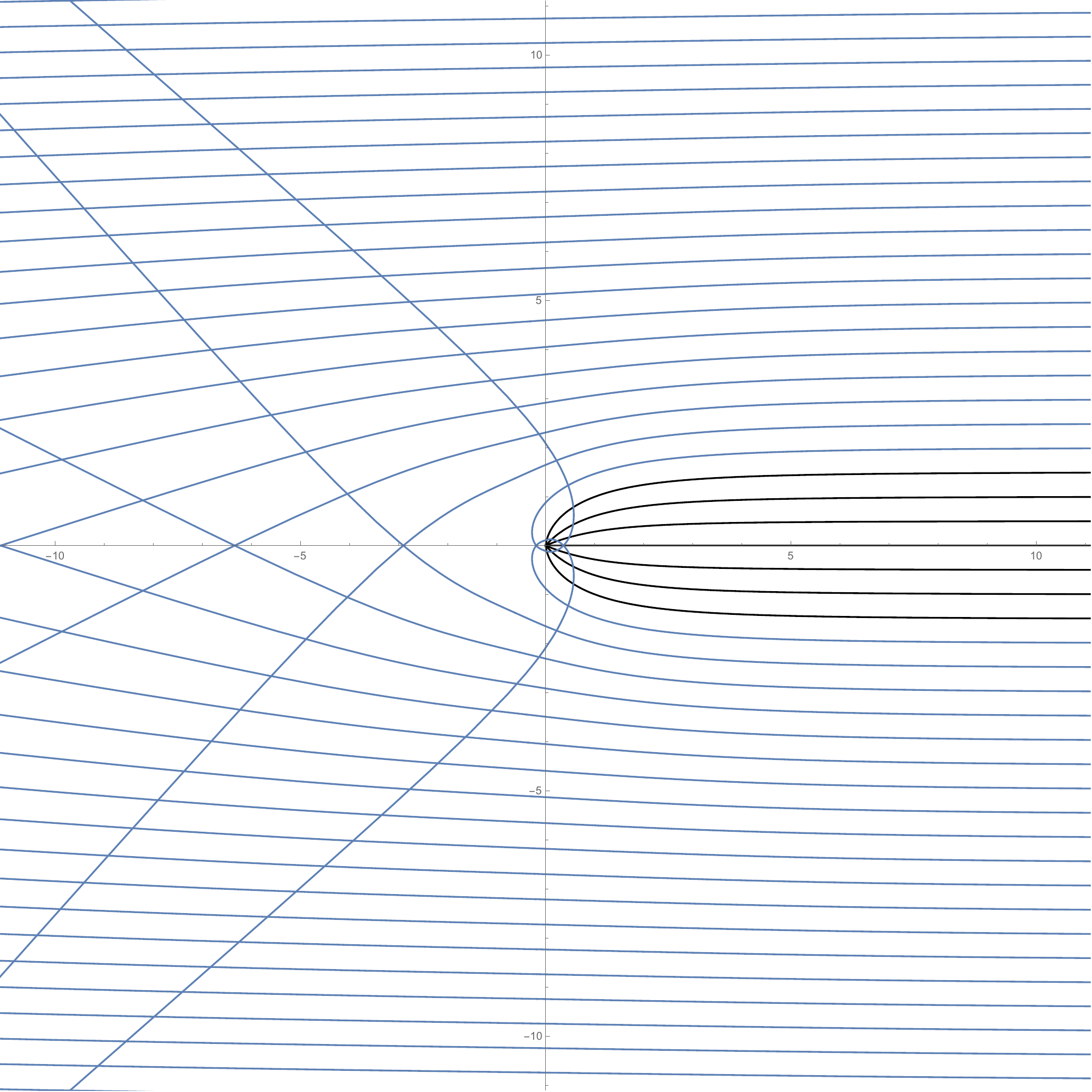}
		\caption*{\scriptsize{$m=0$, $a=2$, $r_{ph}=0$, $|b_{ph}|=2$}}
	\end{minipage}
	\hfill
	\begin{minipage}[h]{0.325\linewidth}
		\includegraphics[width=1\linewidth]{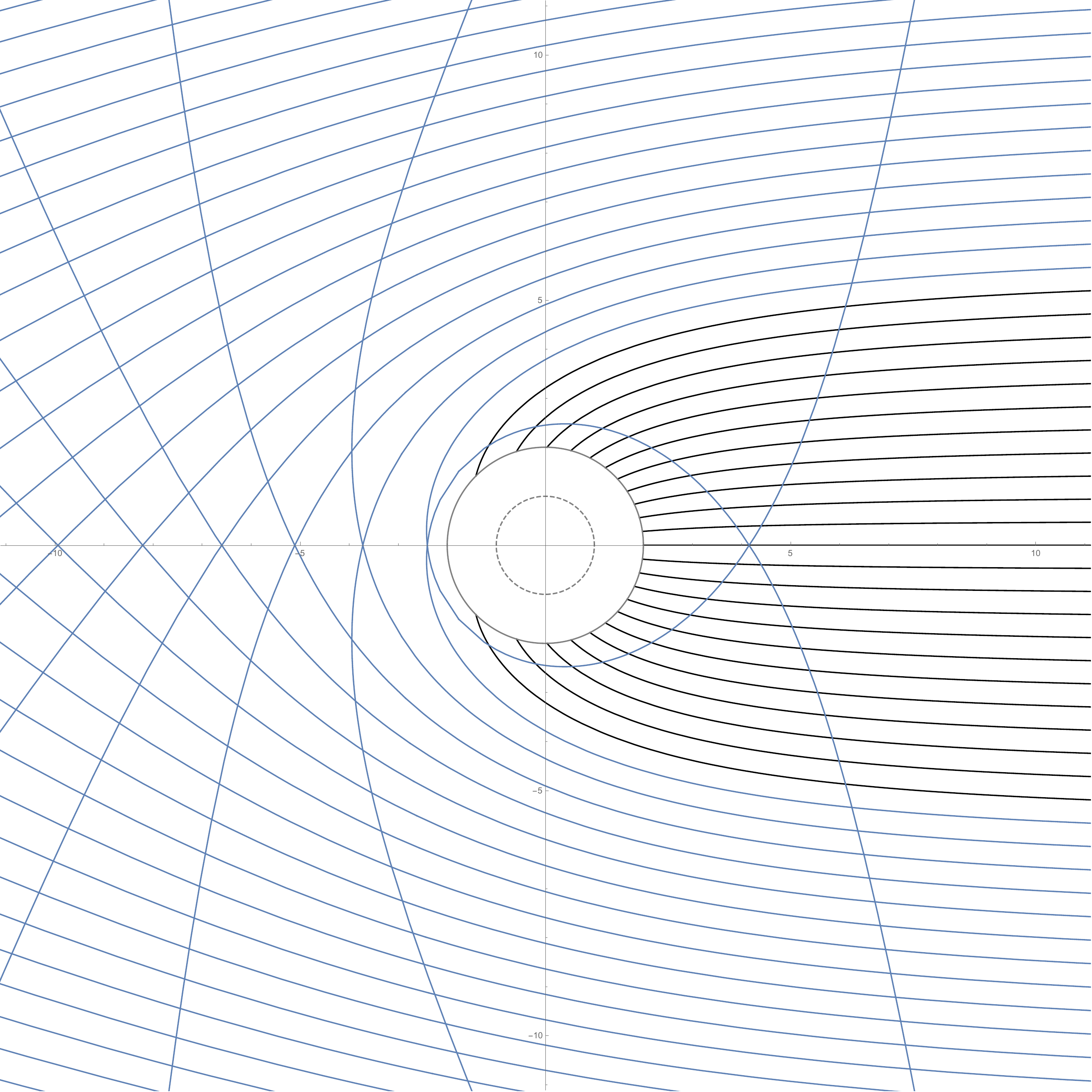}
		\caption*{\scriptsize{$m=1$, $a=2$, $r_{ph}=2$, $|b_{ph}|\approx6.2035$}}
	\end{minipage}
	\hfill
	\begin{minipage}[h]{0.325\linewidth}
		\includegraphics[width=1\linewidth]{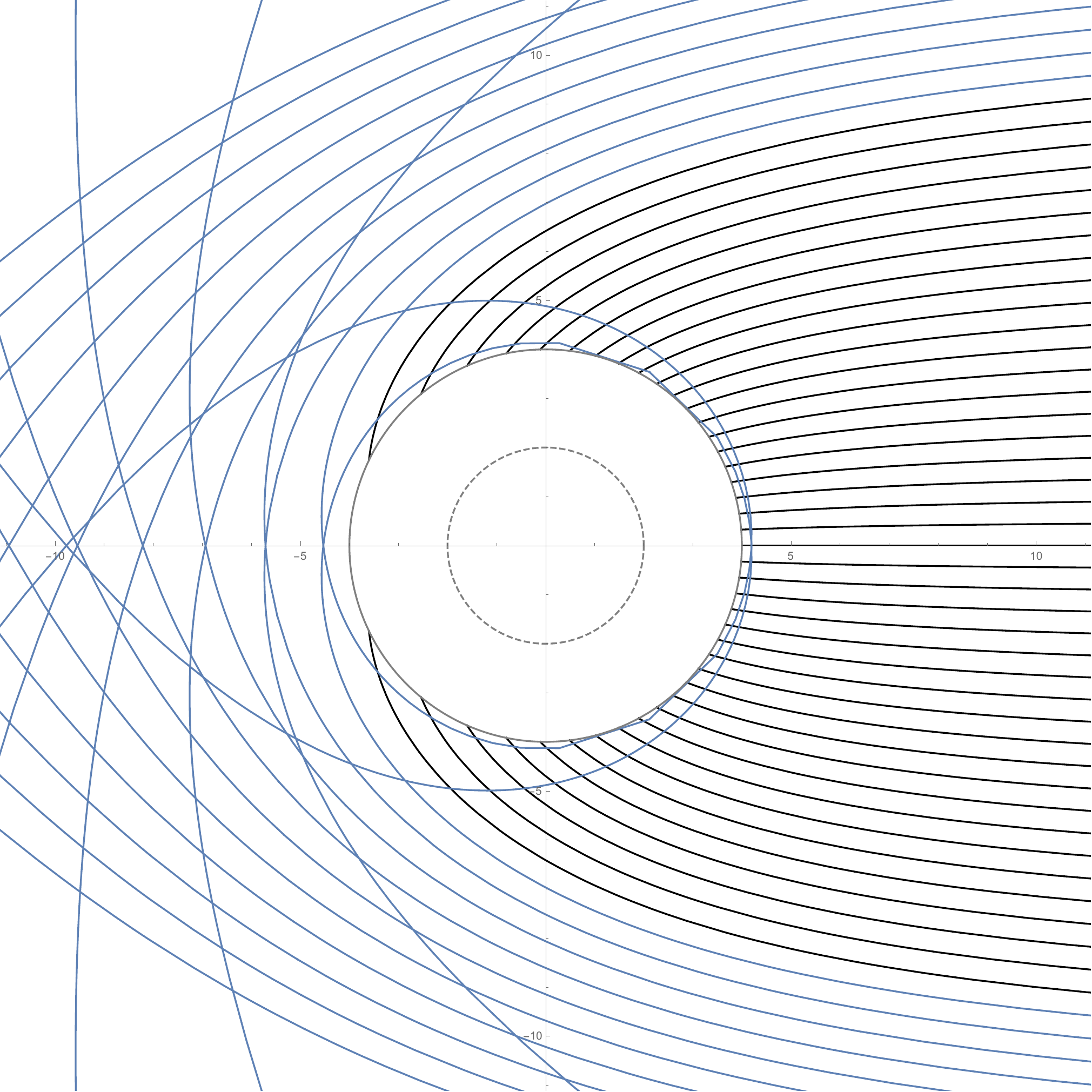}
		\caption*{\scriptsize{$m=2$, $a=2$, $r_{ph}=4$, $|b_{ph}|\approx11.3041$}}
	\end{minipage}
	\caption{\label{AllTraj}Photon trajectories $r(\varphi)$ in the space-time of the Ellis-Bronnikov wormhole ($r>0, \theta=\pi/2$). The black curves are the trajectories of photons passing through the wormhole ($|b|<|b_{ph}|$), the blue curves are the trajectories of photons bending around it ($|b|>|b_{ph}|$). The larger gray circle is the circular orbit of photons $r_{ph}=2m$. The smaller gray circle is the throat $r_{th}=m$.}
\end{figure}

\begin{figure}[t!]
	\begin{minipage}[h]{0.4\linewidth}
		\center{\includegraphics[width=1\linewidth]{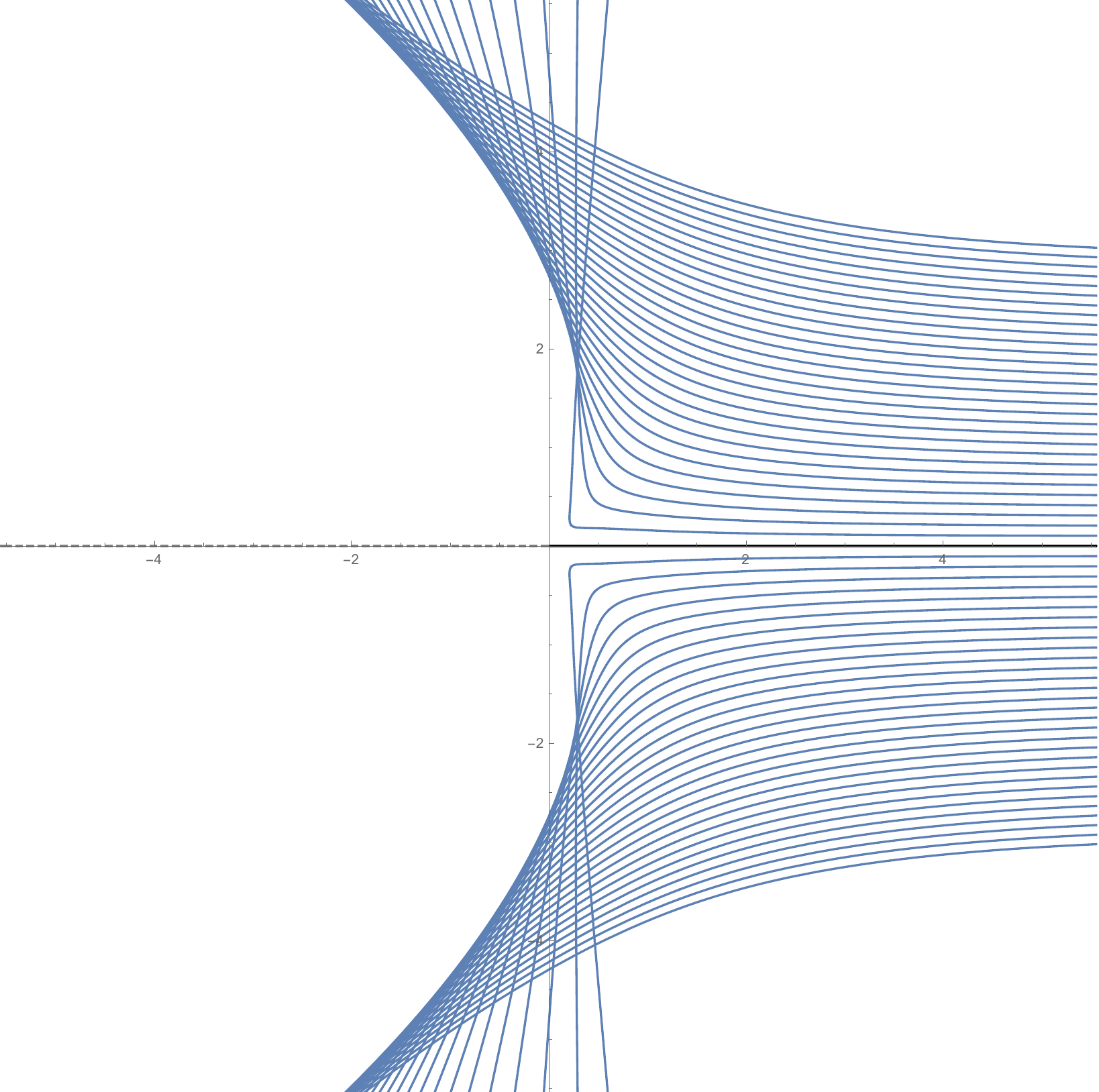}}
		\caption*{\scriptsize{$m=1$, $a=1$, $|b_{ph}|\approx5.6520$}}
	\end{minipage}
	\hspace{1.5cm}
	\begin{minipage}[h]{0.4\linewidth}
		\center{\includegraphics[width=1\linewidth]{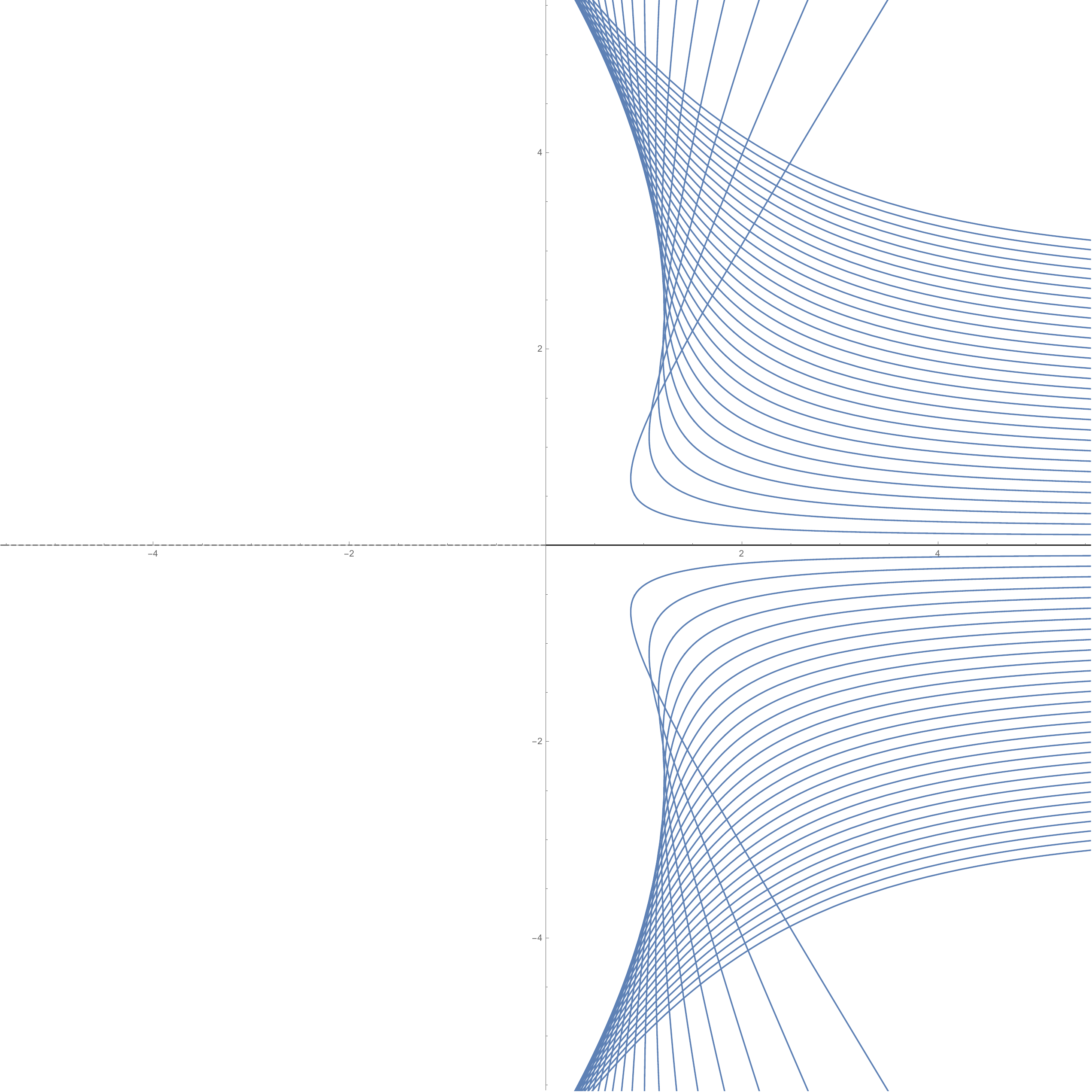}}
		\caption*{\scriptsize{$m=2$, $a=1$, $|b_{ph}|\approx10.9849$}}
	\end{minipage}
	\vfill
	\begin{minipage}[h]{0.4\linewidth}
		\center{\includegraphics[width=1\linewidth]{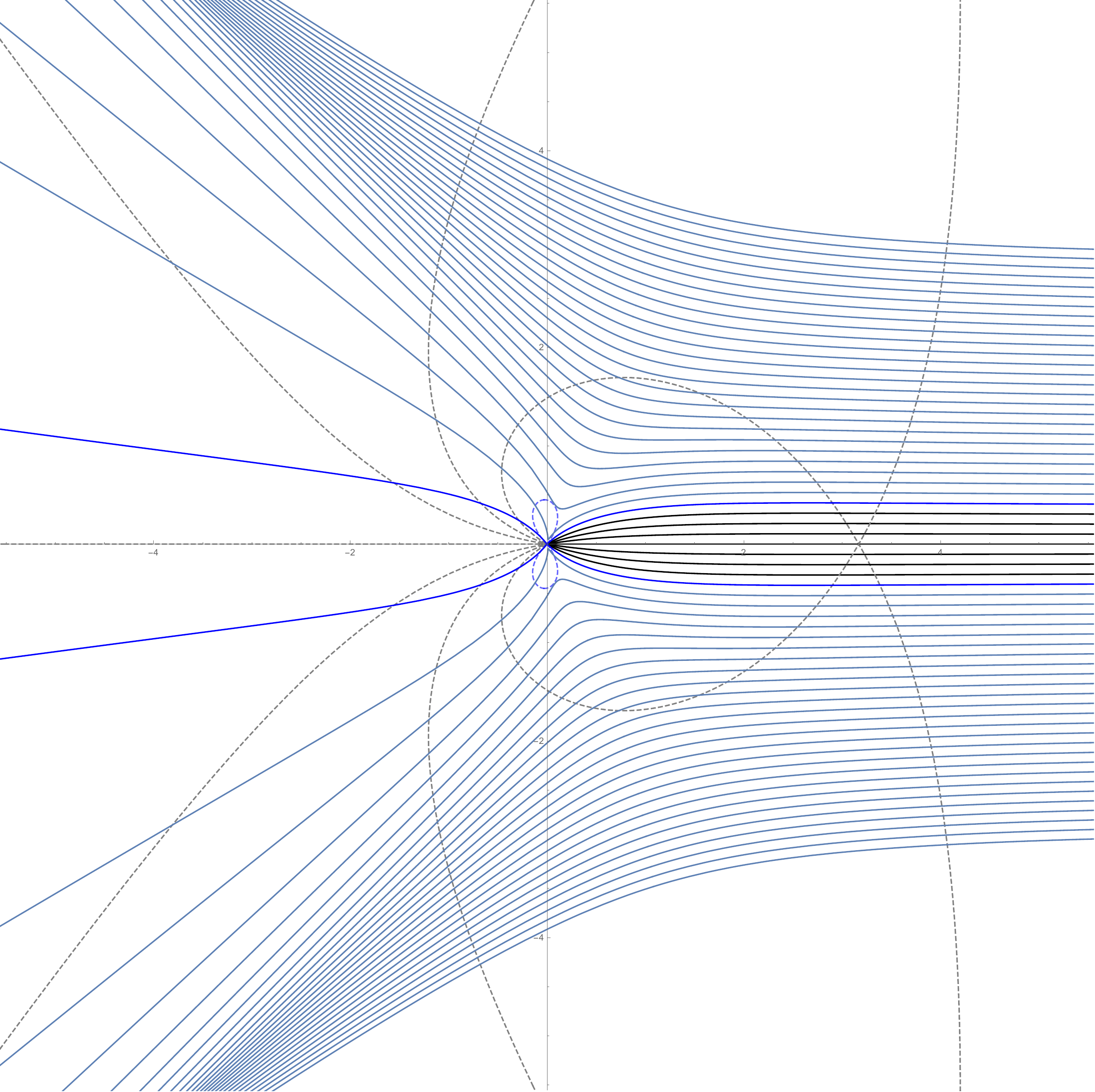}}
		\caption*{\scriptsize{$m=1$, $a=2$, $r_{ph}=2$, $|b_{ph}|\approx6.2035$}}
	\end{minipage}
	\hspace{1.5cm}
	\begin{minipage}[h]{0.4\linewidth}
		\center{\includegraphics[width=1\linewidth]{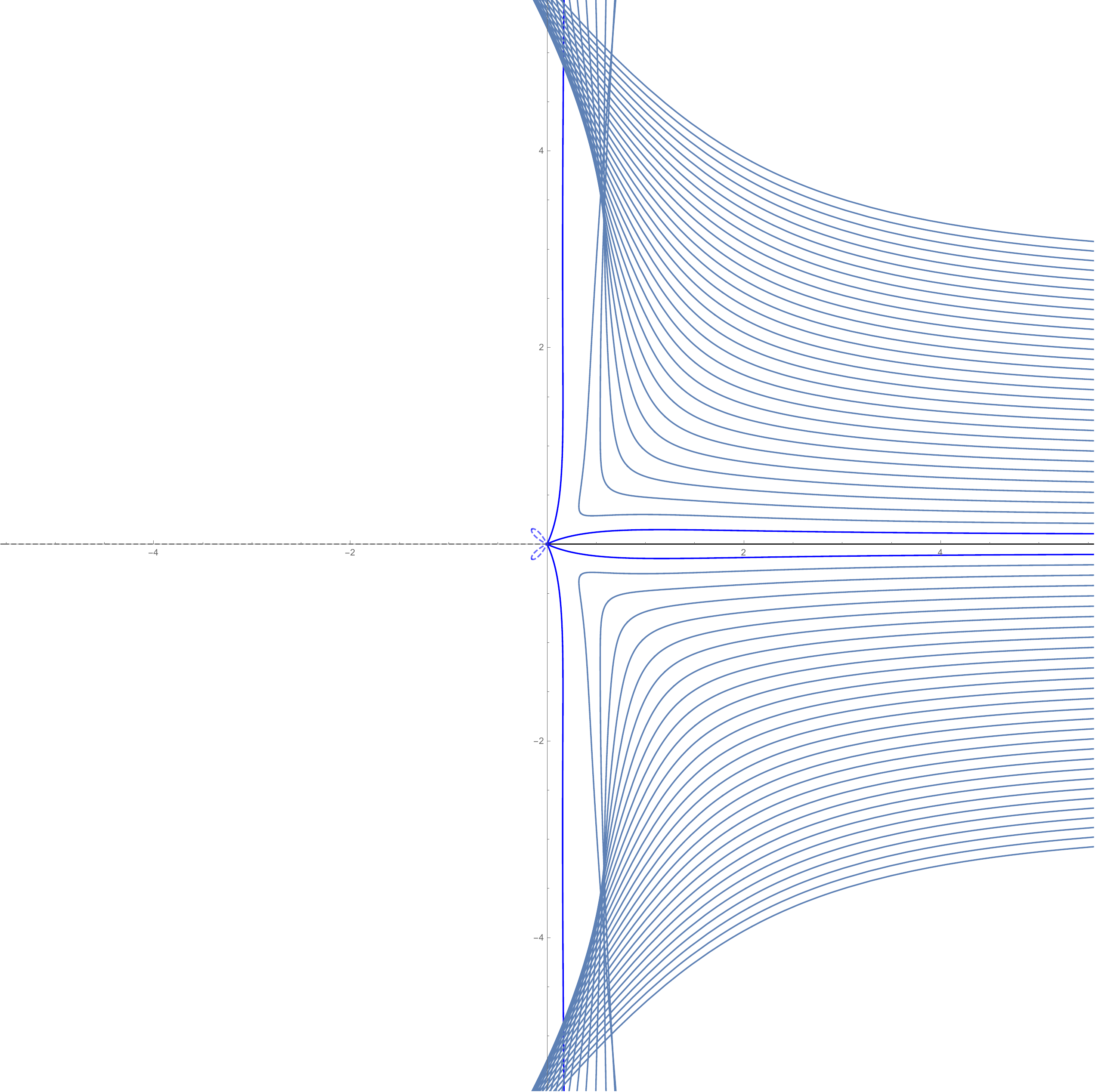}}
		\caption*{\scriptsize{$m=2$, $a=2$, $r_{ph}=4$, $|b_{ph}|\approx11.3041$}}
	\end{minipage}
	\caption{\label{negative}Photon trajectories $r(\phi)$ in the Ellis-Bronnikov wormhole space-time ($r<0, \theta=\pi/2$). Photons with $|b|>|b_{ph}|$ go into the space $r>0$ (black), with $|b|<|b_{ph}|$ go around the wormhole (light blue) or pass into space $r>0$ but return to $r<0$ (blue). The gray curves are photon trajectories in the space $r>0$.}
\end{figure}

Figs. \ref{AllTraj} and \ref{negative} show examples of photon trajectories in the ${\cal R}_+$ and ${\cal R}_-$ regions, respectively. The images were built in such a way that photons start moving at infinity, so that their geodesics are parallel to the line connecting the observer and the wormhole.

In Fig. \ref{AllTraj}, photons fall into the wormhole from infinity in the ${\cal R}_+$ region, where the asymptotic mass of the wormhole is positive. Photons with an impact parameter $|b|>|b_{ph}|$ go around the wormhole (blue curves). Photons with $|b|<|b_{ph}|$ cross the throat of the wormhole and go to the ${\cal R}_-$ region (black curves). If the motion of photons is reversed, then it turns out that photons fall on the wormhole from all sides, go around it and rush to the observer, and only photons with an impact parameter $|b|>|b_{ph}|$ reach his. Thus, photons with $|b|<|b_{ph}|$ form the shadow of the wormhole, and photons with $|b|=|b_{ph}|$ moving in a circular orbit form the boundary of the shadow.

In Fig. \ref{negative}, photons fall into the wormhole from infinity in the ${\cal R}_-$ region, where the asymptotic mass of the wormhole is negative. The picture for photons moving in the ${\cal R}_-$ region is completely different from the picture for photons moving in the ${\cal R}_+$ region. In this region, the wormhole throat plays the role of a repulsive center, scattering photons with an impact parameter $|b|>|b_{ph}|$ and only photons with $|b|<|b_{ph}|$ pass through the neck into the region ${\cal R}_+$. This behavior of light has not been observed in our Universe, so in what follows we will consider only the ${\cal R}_+$ region and assume that the observer is in the ${\cal R}_+$ region.

\onecolumngrid\
\twocolumngrid\

\onecolumngrid\
\twocolumngrid\

\subsection{Particle trajectories and the innermost stable circular orbit (ISCO)}
In this section, we consider some features of the motion of massive particles ($\mu\neq 0$). Since we are interested in a thin accretion disk, we will consider the motion of particles in the equatorial plane of the wormhole ($\theta=\pi/2, Q=0$).

The formation of an accretion disk requires the presence of stable circular orbits of particles, which are determined by the following conditions \cite{BardeenPress}
\begin{equation}
R(r)=0,\quad R'(r)=0, \quad R''(r)\leq 0.
\end{equation}
The first two equations determine the parameters $E, L$ of particles moving in circular orbits:
\begin{eqnarray}
	&&\frac{E}{\mu}=e^{u(r)}\sqrt{\frac{r^2+a^2+e^{2u(r)}(L/\mu)^2}{r^2+a^2}},
	\label{Eisco}\\
	&&\frac{L}{\mu}=e^{-u(r)}\sqrt{\frac{m(r^2+a^2)}{r-2m}}.
	\label{Lisco}
\end{eqnarray} 
It can be seen from the (\ref{Lisco}) equation that circular orbits of particles exist only for $r>2m$. Thus, an accretion disk around a wormhole can form only in the ${\cal R}_+$ region.

The inner boundary of the stable motion of massive particles in the accretion disk is determined by the innermost stable circular orbit $r_{ISCO}$ corresponding to $R''(r)=0$:
\begin{equation}
     r_{ISCO}=3m+\sqrt{a^2+5m^2}.\label{risco2}
\end{equation}

Following the article \cite{Dokuchaev:2019jqq}, in what follows we will consider the inner part of the thin accretion disk $r_{th}<r<r_{ISCO}$, which does not contain stable orbits. The motion of matter in this area is completely non-stationary and depends only on the gravitational field of the wormhole. 

\section{Wormhole observation} \label{Sec_WO}
A wormhole can be observed if there is a radiating substance around it. The action of the gravity of the wormhole leads to the fact that the rays of light are deflected and some of them are captured by the wormhole. Because of this, the observer sees a dark spot in place of the wormhole. The size of the dark spot will depend on where the emitting material is located. If the luminous background is far from the wormhole ($r_s\gg r_{th}$), then the classical wormhole {\em shadow} of the maximum size is observed, which we will simply call the shadow. The boundary of the shadow in this case is formed by photons that move in circular orbits, as shown above. If the emitting material is close to the wormhole, for example, if the wormhole is surrounded by an accretion disk, then the {\em silhouette} of the wormhole throat is observed. In what follows, we will assume that photons emitted in the ${\cal R}_-$ region do not reach the observer in the ${\cal R}_-$ region or experience such a strong gravitational displacement that they cannot be observed. The silhouette of the throat is formed by photons that are emitted at the throat and reach the observer. Since the size of the throat is smaller than the size of the circular orbit of photons, the size of the silhouette of the throat will also be smaller than the size of the wormhole shadow. In this section, we get the size of the wormhole shadow, the size of the throat silhouette, compare them with each other and with the shadow and silhouette of the event horizon of a Schwarzschild black hole.
\subsection{Shadow of the Ellis-Bronnikov wormhole}
In this section, we obtain the size of the classical shadow that is observed if the glowing background is far from the wormhole ($r_s\gg r_{th}$).

To obtain the size of the shadow that the observer sees, it is necessary to obtain the coordinates of the incoming light beam in the observer's sky. They can be found in the following way \cite{Vasquez}:
\begin{equation}\label{skycoor}
	\alpha=-r_O^2\sin\theta_O\left.\frac{d\varphi}{dr}\right|_{r_O}, \quad \beta=r_O^2\left.\frac{d\theta}{dr}\right|_{r_O},
\end{equation}
where $r_O$ is a distance between the wormhole and observer, $\theta_O$ is an angle between the wormhole axis of rotation and a line connecting the observer and wormhole.

As we have shown above, the boundary of the wormhole shadow is formed by photons that move in circular orbits. Then, assuming $r_O\gg m$ and $\theta_O = \pi/2$ and taking into account spherical symmetry, it is easy to show that the shadow boundary is appeared as a circle with the radius $\alpha$:
\begin{equation}
	\left.\alpha\right|_{r\to+\infty}=|b_{ph}|.\label{shadow}
 \end{equation}
\begin{figure}[h!]
	\noindent\centering
	{\includegraphics [width=1.0\linewidth]{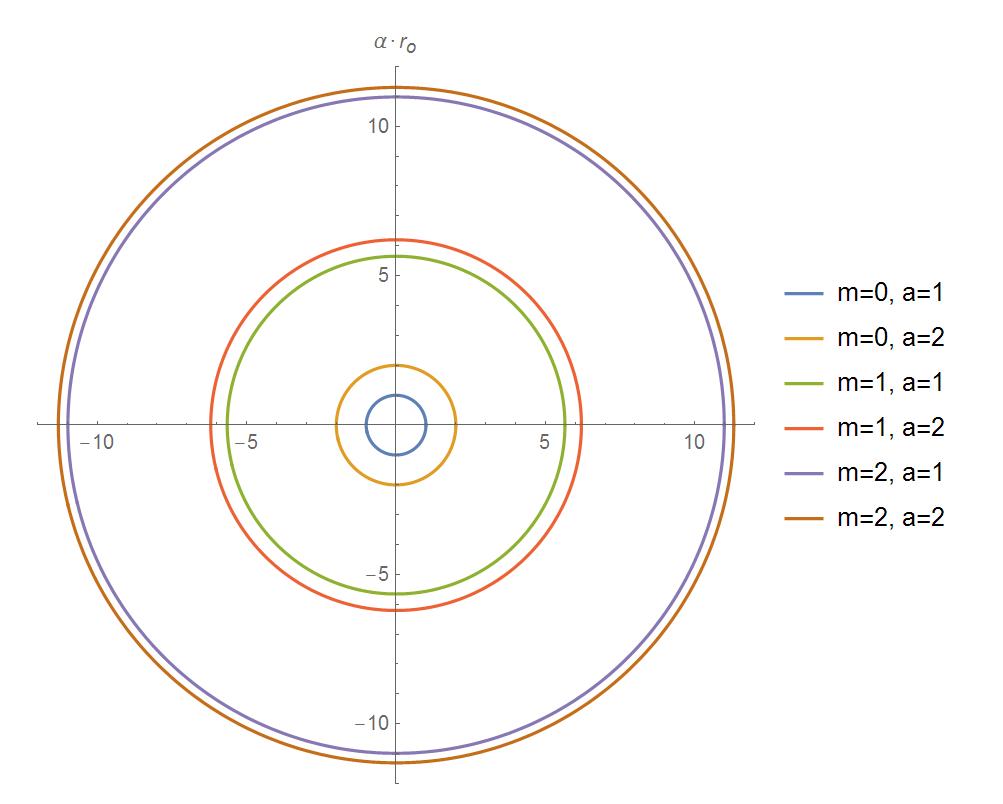}}
	\caption{\label{shadowfig}The size of the Ellis-Bronnikov wormhole shadow for various parameters $m$ and $a$ in the ${\cal R}_+$ region.}
\end{figure}

\subsection{Silhouette of the Ellis-Bronnikov wormhole}
At present, observations of classical wormhole shadows are improbable, since the intensity of the radiation from the distant background is too low to be detected by modern telescopes. Therefore, it is much more likely to observe the silhouette of the throat of an accreting wormhole.
In this section, we obtain an equation for the size of the silhouette of the Ellis-Bronnikov wormhole throat, following the article \cite{Dokuchaev:2019jqq}.

The parameters of photons that move along a given trajectory can be found from the integral equations of motion:
\begin{equation}
\int\frac{e^{2u(r)}dr}{(r^2+a^2)\sqrt{R(r)}}=\int\frac{d\theta}{\sqrt{\Theta(\theta)}},
\end{equation}
in which the integrals are taken along the trajectory.
According to the Cunningham-Bardeen classification scheme of multiple lensed images \cite{Cunningham1, Cunningham2}, photons forming the direct image of the source do not cross the wormhole equatorial plane along the whole path from the emitter to the observer. For such photons, the integral equations of motion are as follows
\begin{equation}
\int_{r_s}^{r_O}\frac{e^{2u(r)}dr}{(r^2+a^2)\sqrt{R(r)}}=
\int_{\theta_O}^{\theta_s}\frac{d\theta}{\sqrt{\Theta(\theta)}}
\label{noturnpoint}
\end{equation}
\begin{equation}
	=\int_{\theta_{min}}^{\theta_s}\frac{d\theta}{\sqrt{\Theta(\theta)}}+\int_{\theta_{min}}^{\theta_O}\frac{d\theta}{\sqrt{\Theta(\theta)}},
\label{turnpoint}
\end{equation}
where $\theta_{min}=\mathop{\rm arctan}\left(q/\sqrt{q^2+l^2}\right)$ is the turning point in the polar $\theta$-direction, determined from the equation $\Theta(\theta) = 0$.

In the simplest case, when the observer and the wormhole are in the $\theta=\pi/2$ plane, the boundary of the throat image visible to a distant observer is given by the solution of the integral equation
\begin{equation}\label{silint}
\int_{m}^{\infty}\frac{e^{2u(r)}dr}{(r^2+a^2)\sqrt{R(r)}}=2\int_{\theta_{min}}^{\pi/2}\frac{d\theta}{\sqrt{\Theta(\theta)}}.
\end{equation}
From equations (\ref{skycoor}) assuming $|r_O|\gg m$ and $\theta_O = \pi/2$, we obtain the silhouette radius of the wormhole throat
\begin{equation}
\alpha_{th}=\sqrt{q^2+l^2}.\label{throat}
\end{equation}
Thus, taking the integral on the right side of (\ref{silint}), we obtain the final equation for the radius of the throat silhouette:
\begin{equation}
    \int_m^{\infty}\frac{e^{\frac{2m}{a}\left(\arctan\frac{r}{a}-\frac{\pi}{2}\right)}dr}{(r^2+a^2)\sqrt{1-e^{\frac{4m}{a}\left(\arctan\frac{r}{a}-\frac{\pi}{2}\right)}\frac{\alpha_{th}^2}{r^2+a^2}}}=\frac{\pi}{\alpha_{th}}.\label{sileq}
\end{equation}
This equation is transcendental, and its solution can be obtained numerically.
\subsection{Comparison of the sizes of the Ellis-Bronnikov wormhole shadow and the Schwarzschild black hole shadow}
To understand whether it is possible to distinguish a wormhole from a black hole by the size of the observed shadow, we compare the results obtained for the Ellis-Bronnikov wormhole with the radii of the shadow and silhouette of the event horizon of a Schwarzschild black hole. 

\begin{figure}[h!]
	\noindent\centering
	\includegraphics[width=0.8\linewidth]{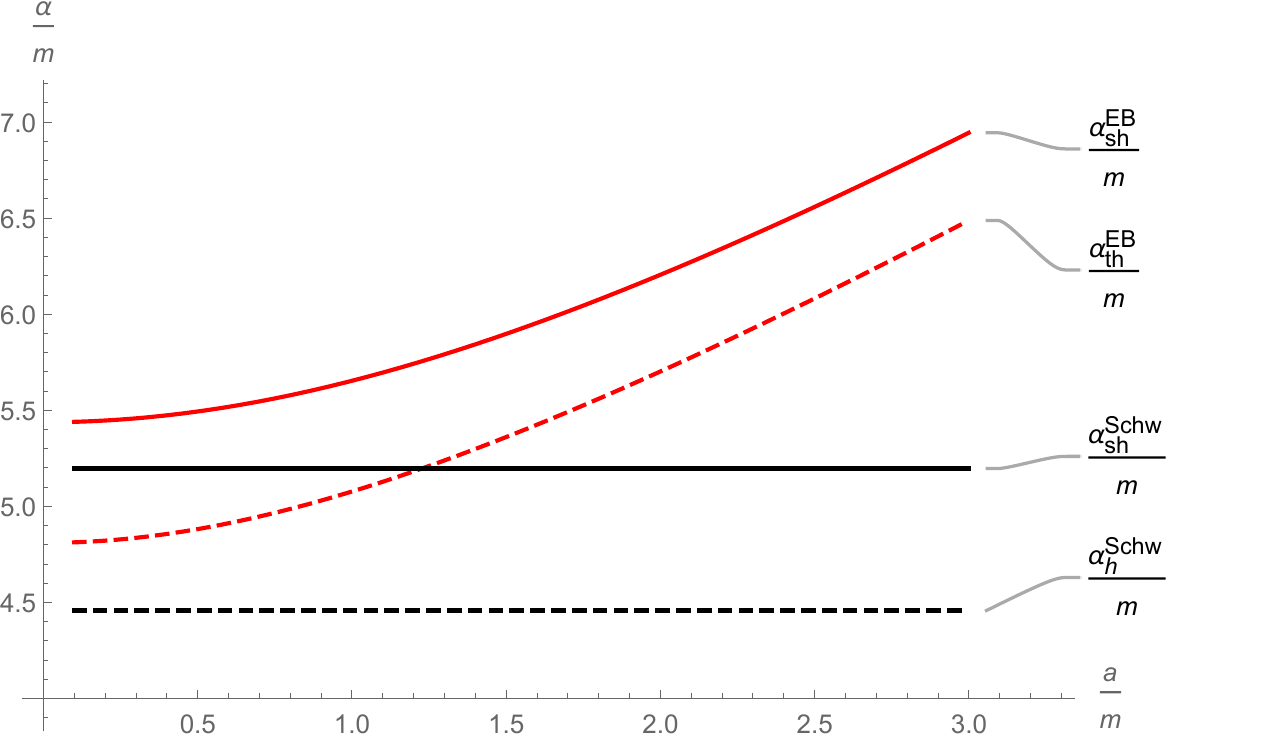}
	\caption{\label{radius}Dependence of the radii of the shadow $\alpha_{sh}^{EB}$ (solid red line) and the neck silhouette $\alpha_{th}^{EB}$ (red dotted line) of the Ellis-Bronnikov wormhole on the $a$ parameter. The plot also shows the radii of the shadow $\alpha_{sh}^{Schw}$ (black solid line) and the silhouette of the event horizon $\alpha_{h}^{Schw}$ (dashed black line) of the Schwarzschild black hole.}
\end{figure}

In Ref. \cite{Dokuchaev:2018ibr} it was shown that
\begin{equation}
 \alpha_{sh}^{Schw}=3\sqrt{3}m\approx 5.196m,   \quad \alpha_{h}^{Schw}\approx 4.457m.
\end{equation}
Numerically solving the equations (\ref{shadow}) and (\ref{sileq}), we constructed the dependence of radii of the shadow $\alpha_{sh}^{EB}$ and the throat silhouette $\alpha_{th}^{EB}$ of the Ellis-Bronnikov wormhole on the $a$ parameter (Fig. \ref{radius}).
Fig. \ref{radius} shows that for any throat parameter $a$, the shadow and silhouette of the Ellis-Bronnikov wormhole throat are larger than the shadow and silhouette of the event horizon of the Schwarzschild black hole, respectively:
\begin{equation}
    \alpha_{sh}^{EB}>\alpha_{sh}^{Schw}, \quad \alpha_{th}^{EB}>\alpha_{h}^{Schw}.
\end{equation}



\section{Image of an accreting Ellis-Bronnikov wormhole}
In this chapter, we obtain an image of an accreting Ellis-Bronnikov wormhole by following \cite{Dokuchaev:2019jqq}. We use the model of a geometrically thin optically transparent accretion disk. In this model, we consider the inner part of the accretion disk $r_{th}<r<r_{ISCO}$, in which disk fragments move along geodesics with parameters $E$ and $L$ from equations (\ref{Eisco}, \ref{Lisco}) corresponding to $r=r_{ ISCO}$. In addition, when calculating the observed radiation, we assume that the energy flux in the comoving system of fragments is isotropic and persists until they reach the throat of $R_{th}$.

\subsection{Locally nonrotating frames}
In any stationary spherically symmetric asymptotically flat spacetime, one can introduce locally nonrotating reference frames (LNRFs) \cite{BardeenPress} in which observers are moving along the world lines $r = const, \theta = const, \varphi = const$.  The orthonormal tetrad carried by such an observer (the set of LNRF basis vectors) at the point $t, r, \theta, \varphi$ for the Ellis-Bronnikov metric (\ref{BEmetric}) is given by 
\begin{eqnarray}
	\mathbf{e}_{(t)}&=&
	e^{-u(r)}\frac{\partial}{\partial t},
	\label{LNRFlow}\\
    \mathbf{e}_{(r)}&=&
    e^{u(r)}\frac{\partial}{\partial r},
    \\
    \mathbf{e}_{(\theta)}&=&
    \frac{e^{u(r)}}{\sqrt{r^2+a^2}}\frac{\partial}{\partial \theta},
    \\
	\mathbf{e}_{(\varphi)}&=&
	\frac{e^{u(r)}}{\sin\theta\sqrt{r^2+a^2}}\frac{\partial}{\partial \varphi}.
\end{eqnarray} 
The corresponding basis of one-forms (or covariant basis vectors) is
\begin{eqnarray}
	\mathbf{e}^{(t)}&=&
	e^{u(r)}\mathbf{d}t,\label{tLNRFup}
	\\
    \mathbf{e}^{(r)}&=&
    e^{-u(r)}\mathbf{d}r,
    \\
    \mathbf{e}^{(\theta)}&=&
    e^{-u(r)}\sqrt{r^2+a^2}\mathbf{d}\theta,
    \\
	\mathbf{e}^{(\varphi)}&=&
	e^{-u(r)}\sin\theta\sqrt{r^2+a^2}\mathbf{d}\varphi.\label{LNRFup}
\end{eqnarray} 
Equations (\ref{LNRFlow} -- \ref{LNRFup}) define the components of basis vectors in the LNRF:
\begin{equation}
    \mathbf{e}_{(\nu)}=e^{i}_{(\nu)}\frac{\partial}{\partial x^{i}}, \quad \mathbf{e}^{(\nu)}=e_i^{(\nu)}\mathbf{d}x^{i}.
\end{equation}

\subsection{Gravitational redshift and Doppler effect}
Photons emitted by the matter of the accretion disk and reaching the observer experience a gravitational redshift and a Doppler shift. To calculate the energy of a photon registered by an observer, it is convenient to use the locally non-rotating frame described above.

In general, the 4-velocity $\mathbf{u}$ has the LNRF components \cite{BardeenPress}
\begin{equation}
    u^{(\nu)}=u^je_j^{(\nu)}
\end{equation}
where the $u^j$ come from equations (\ref{ut}--\ref{uphi}), and the $e_j^{(\nu)}$ from equations (\ref{tLNRFup}--\ref{LNRFup}). The 3-velocity relative to the LNRF has components
\begin{equation}
    v^{(\mu)}=\frac{u^je_j^{(\mu)}}{u^ie_i^{(t)}}, \quad \mu=r, \theta, \phi.
\end{equation}
In particular, the azimuthal and radial components of the velocity of a small accretion disk fragment at a radius $r$ with orbital parameters $E$, $L$ and $Q = 0$ is
\begin{eqnarray}
    v^{(\varphi)}&=& \frac{L e^{2 u(r)}}{E \sqrt{a^2+r^2}},\\
    v^{(r)}&=& \frac{\sqrt{-\frac{L^2 e^{4u(r)}}{a^2+r^2}+E^2-\mu^2e^{2u(r)}}}{E}.
\end{eqnarray}
We also need expressions for the components of the photon 4-momentum in the LNRF:
\begin{eqnarray}
    p^{(\varphi)}&=& \frac{l e^u(r)}{\sqrt{a^2+r^2}},\\
    p^{(r)}&=& e^{-u(r)} \sqrt{1-\frac{e^{4 u(r)}\left(l^2+q^2\right)}{a^2+r^2}},\\
    p^{(t)}&=& e^{-u(r)}.
\end{eqnarray}

Consider a fragment of an accretion disk moving with an azimuthal velocity $v^{(\varphi)}$ and a radial velocity $v^{(r)}$ relative to the LNRF. Then the photon energy in the comoving frame of this fragment is \cite{Dokuchaev:2019jqq}
\begin{equation}
    \varepsilon(l,q)=\frac{p^{(t)}-v^{(\varphi)}p^{(\phi)}-v^{(r)}p^{(r)}}{\sqrt{1-\left(v^{(\varphi)}\right)^2-\left(v^{(r)}\right)^2}}.\label{energy}
\end{equation}
The corresponding photon energy shift (ratio of the photon frequency detected by a remote observer to the frequency of the same photon in the comoving frame of the fragment) is $g(l,q)=1/\varepsilon(l,q)$.
\begin{widetext}
\onecolumngrid\
\begin{figure}[t!]
	\begin{minipage}[h]{0.45\linewidth}
		\center{\includegraphics[width=1\linewidth]{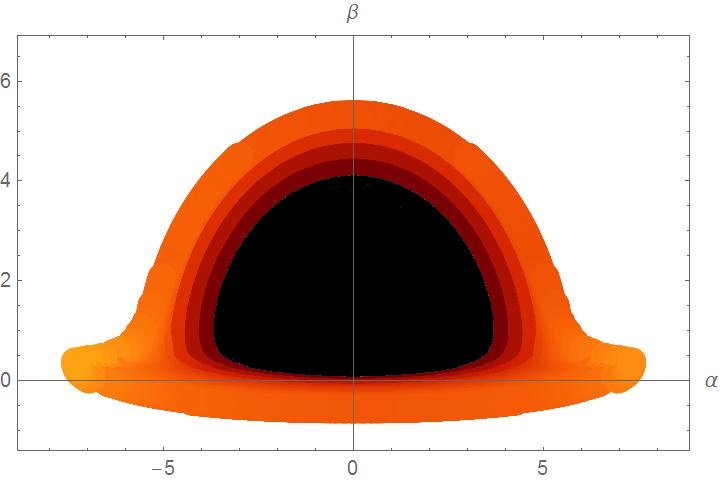}}
	\end{minipage}
	\hfill
	\begin{minipage}[h]{0.45\linewidth}
		\center{\includegraphics[width=1\linewidth]{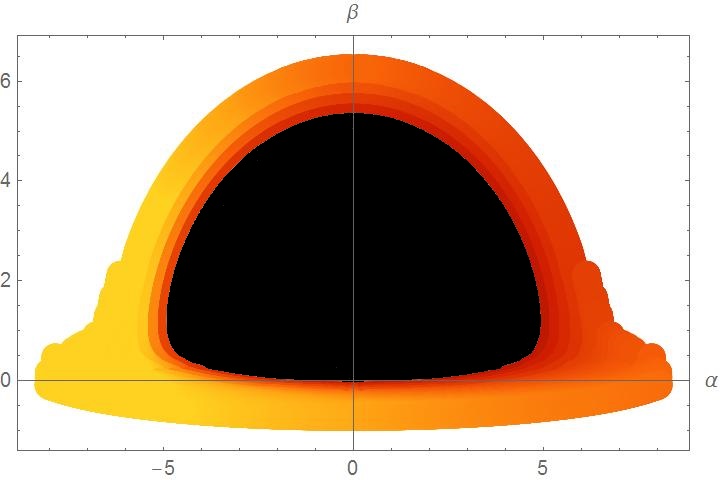}}
	\end{minipage}
	\hfill
	\begin{minipage}[h]{0.05\linewidth}
		\center{\includegraphics[width=1\linewidth]{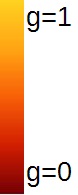}}
	\end{minipage}
	\caption{\label{images} Images of an accreting Schwarzschild black hole (left) and an accreting Ellis-Bronnikov wormhole (right). The observer is at $r_O=10000m, \theta_O=84.24^{\cdot}$}
\end{figure}
\twocolumngrid\
\end {widetext}

\subsection{Comparison}
We build the image, taking into account photons that are emitted in the interior of the accretion disk and have no more than one turning point. The photon parameters $l$ and $q$ are found from the equations (\ref{noturnpoint}), (\ref{turnpoint}) at $r_{th}\leq r_s\leq r_{ISCO}$. The energy shift of an incoming photon is calculated by the formula $g(l, q)=1/\varepsilon(l,q)$, where $\varepsilon(l,q)$ is found from (\ref{energy}).

The Fig. \ref{images} shows images of an accreting Schwarzschild black hole and an accreting Ellis-Bronnikov wormhole. The masses of the objects are taken to be equal, $m_{Schw}=m_{EB}=1$, and the throat parameter of the wormhole is $a=2$. Under these parameters, the area of the dark spot in the Ellis-Bronnikov wormhole image is $38\%$ larger than the area of the dark spot in the Schwarzschild black hole image. With different throat parameters, $a$, the dark spot in the image of a wormhole will also be larger because it is the silhouette of the throat (Fig. \ref{radius}). As for the energy shift, photons emitted near the Ellis-Bronnikov wormhole experience less redshift compared to photons emitted near the Schwarzschild black hole. Moreover, for photons emitted in the innermost stable circular orbit of particles in the spacetime of the Ellis-Bronnikov wormhole, the Doppler effect plays a major role in the energy shift.

\section{Summary}
In this paper we have explored gravitational lensing properties of the Ellis-Bronnikov wormhole spacetime with the metric (\ref{BEmetric}),
which is a generalization of the more simple Ellis solution (\ref{Ellis_wh}). The Ellis-Bronnikov wormhole is characterized by two parameters $a$ and $m$, related to a wormhole throat radius and an asymtotic wormhole mass, respectively, while the Ellis wormhole is massless, i.e. $m=0$. In our work, we have investigated in details the propagation of light in the Ellis-Bronnikov wormhole spacetime, the formation of its shadow and silhouette, and the formation of an image of matter disk accreting around the wormhole. As the result, we have obtained an analytical dependence of geometrical and physical characteristics of various images of the Ellis-Bronnikov wormhole on the parameters $a$ and $m$ and constructed their graphical representations. Then, all images obtained for the Ellis-Bronnikov wormhole have been compared with those for the Schwarzschild black hole.
Summarizing the results of investigation, we can conclude that there exists the rather strong difference between images of a wormhole and a black hole. This difference can be used in future astrophysical observations to distinguish black holes and wormholes.

\section*{Acknowledgments}
This work is supported by the RSF grant No. 21-12-00130 and partially carried out in accordance with the Strategic Academic Leadership Program "Priority 2030" of the Kazan Federal University.


\end{document}